\documentclass[journal=nalefd,manuscript=letter]{achemso}

\usepackage[version=3]{mhchem} % Formula subscripts using \ce{}

\author{F. Chiaruttini}
\author{T. Guillet}
\author{C. Brimont}
\author{B. Jouault}
\author{P. Lefebvre}
\affiliation[UM]
{L2C, University of Montpellier, CNRS, Montpellier, France}
\author{S. Chenot}
\affiliation[CHREA]
{CRHEA, Universit\'e C\^ote d’Azur, CNRS, Valbonne, France}
\author{Y. Cordier}
\affiliation[CHREA]
{CRHEA, Universit\'e C\^ote d’Azur, CNRS, Valbonne, France}
\author{B. Damilano}
\affiliation[CHREA]
{CRHEA, Universit\'e C\^ote d’Azur, CNRS, Valbonne, France}
\author{M. Vladimirova}
\email{maria.vladimirova@umontpellier.fr}
\affiliation[UM]
{L2C, University of Montpellier, CNRS, Montpellier, France}
\title[Trapping dipolar exciton fluids in GaN/(AlGa)N nanostructures]
  {Trapping dipolar exciton fluids in GaN/(AlGa)N nanostructures
}

\keywords{exciton fluid, electrostatic traps, cooling, gallium nitride}

\begin{document}
%\begin{tocentry}
%
%Some journals require a graphical entry for the Table of Contents.
%This should be laid out ``print ready'' so that the sizing of the
%Inside the \texttt{tocentry} environment, the font used is Helvetica
%8\,pt, as required by \emph{Journal of the American Chemical
%The surrounding frame is 9\,cm by 3.5\,cm, which is the maximum
%permitted for  \emph{Journal of the American Chemical Society}
%graphical table of content entries. The box will not resize if the
%content is too big: instead it will overflow the edge of the box.
%
%This box and the associated title will always be printed on a
%separate page at the end of the document.
%
%\end{tocentry}

\begin{abstract}%
Dipolar excitons offer a rich playground for both design of novel optoelectronic devices and fundamental many-body physics.
Wide GaN/(AlGa)N quantum wells 
host a new and promising realization of dipolar excitons. 
We demonstrate the in-plane confinement and cooling of these excitons, when trapped in the electrostatic potential created by semitransparent electrodes 
of various shapes deposited on the sample surface.
This result is a prerequisite for the electrical control of the exciton densities and fluxes, as well for studies of the complex phase diagram of these dipolar bosons at low temperature. 
%
%Up to $500$~$\mu$m$^2$ area homogeneously filled with thermalized exciton fluid at 
%up to Mott density is  
 %
%
%
%  This is an example document for the \textsf{achemso} document
%  class, intended for submissions to the American Chemical Society
%  for publication. The class is based on the standard \LaTeXe\
%  \textsf{report} file, and does not seek to reproduce the appearance
%  of a published paper.
%
%  This is an abstract for the \textsf{achemso} document class
%  demonstration document.  An abstract is only allowed for certain
%  manuscript types.  The selection of \texttt{journal} and
%  \texttt{manuscript} will determine if an abstract is valid.  If
%  not, the class will issue an appropriate error.
\end{abstract}

%\section{Introduction}

Dipolar excitons, Coulomb-bound but spatially separated electron-hole pairs, have a long life-time and a built-in dipole moment that  offer an opportunity for the cooling and electrical control of exciton fluids \cite{LozovikYudson,Miller1985,Huber1998,Ivanov1999,Butov1999,High2008,Butov2017}. 
Various intriguing quantum phenomena including Bose-Einstein-like condensation, darkening and superfluidity of excitons have been recently reported \cite{High2012,Shilo2013,Schinner2013,Cohen2016,Butov2016,Combescot2017,Anankine2017,Misra2018}. 
Albeit demonstrated at very low temperatures, those phenomena are promising for better understanding of new states of matter, but also for potential applications in excitonic devices with novel functionalities \cite{Butov2017}. 

The recent emergence of high quality wide-bandgap semiconductor quantum wells  (QWs) and two-dimensional Van der Waals heterostructures, hosting dipolar excitons with large exciton binding energies and built-in electric fields, has given a new impetus to this research \cite{Kuznetsova2015,Fedichkin2016,Rivera2015,Ross2017,Unuchek2018}.  
Room temperature exciton transport in GaN/(AlGa)N  QWs has been demonstrated \cite{Fedichkin2016}, as well as its electrical control in MoS$_2-$WSe$_2$ heterostructures \cite{Unuchek2018}.
The latter is particularly important in the quest for exciton-based efficient interconnects between optical data transmission and electrical processing systems \cite{Butov2017}.

However, as demonstrated in GaAs-based QWs, trapping, cooling and control of the density of dipolar excitons is essential to address various collective states, such as Bose-Einstein-like condensates, that may form at low temperatures \cite{Butov2016,Combescot2017}. 
While excitons in GaN-based  nanostructures have relatively high binding energies and small Bohr radii \cite{Gil2014} (as compared to GaAs-based structures) and thus could exhibit quantum properties at higher temperatures \cite{Laikhtman2009}, the additional challenges that must be faced to achieve such  degree of control are numerous.
These may include nonradiative losses \cite{Shockley1952}, megavolt per centimeter-strong built-in electric fields \cite{Grandjean1999}, an exponential dependence of the lifetime on the exciton density under high excitation conditions\cite{Lefebvre2004,Liu2016,Chen2018}, and the guided-and-scattered light that must be distinguished from the exciton photoluminescence (PL) \cite{Fedichkin2015,Fedichkin2016}.

%Pioneering studies realised in atomic physics have shown that this strategy provides a deterministic route to manipulate quantum states of matter. Electrostatic trapping of dipolar excitons in GaAs coupled QWs allowed for observation of Bose-Einstein condensation and superfluidity.
%
%The objective of this work is a demonstration of the macroscopic-size cold exciton fluid, electrostatically trapped in the plane of  GaN/(GaAl)N QW, in order to path the way towards realisation  of robust quantum fluids of dipolar excitons.
%
%In GaN QWs excitons are created by optical 
%
%Because strong dipole-dipole coupling between excitons leads to fast expansion of the optically created hot exciton cloud,  trapping of the excitons is mandatory to create homogeneous exciton density in an macroscopic-size  surface area. 
%In the structures based on GaN, the chalenges that must be faced to achieve these goals include non-radiative recombination, megavolt-strong built-in electric fields, an exponential dependence of the lifetime on the exciton density, and high density of the scattered light that must be distinguished from the exciton PL . 

In this letter we overcome those challenges and report on the realization of  $\sim
10$~$\mu$m $\times50$~$\mu$m-size thermalized exciton fluid, trapped in the plane of  a GaN/(GaAl)N quantum well grown on a free-standing GaN substrate. 
Our objective is to obtain an area uniformly filled with excitons, with a density that can be controlled by the excitation power.
To this end we fabricate electrostatic traps of various shapes by depositing semi-transparent metallic layers on top of the semiconductor nanostructure. 
Such strategy has already been used in GaAs-based heterostructures \cite{Huber1998,Hammack2006},  but here it has to be adapted to the case of strongly polar heterostructures.

\begin{figure}
\includegraphics [width=1\columnwidth] {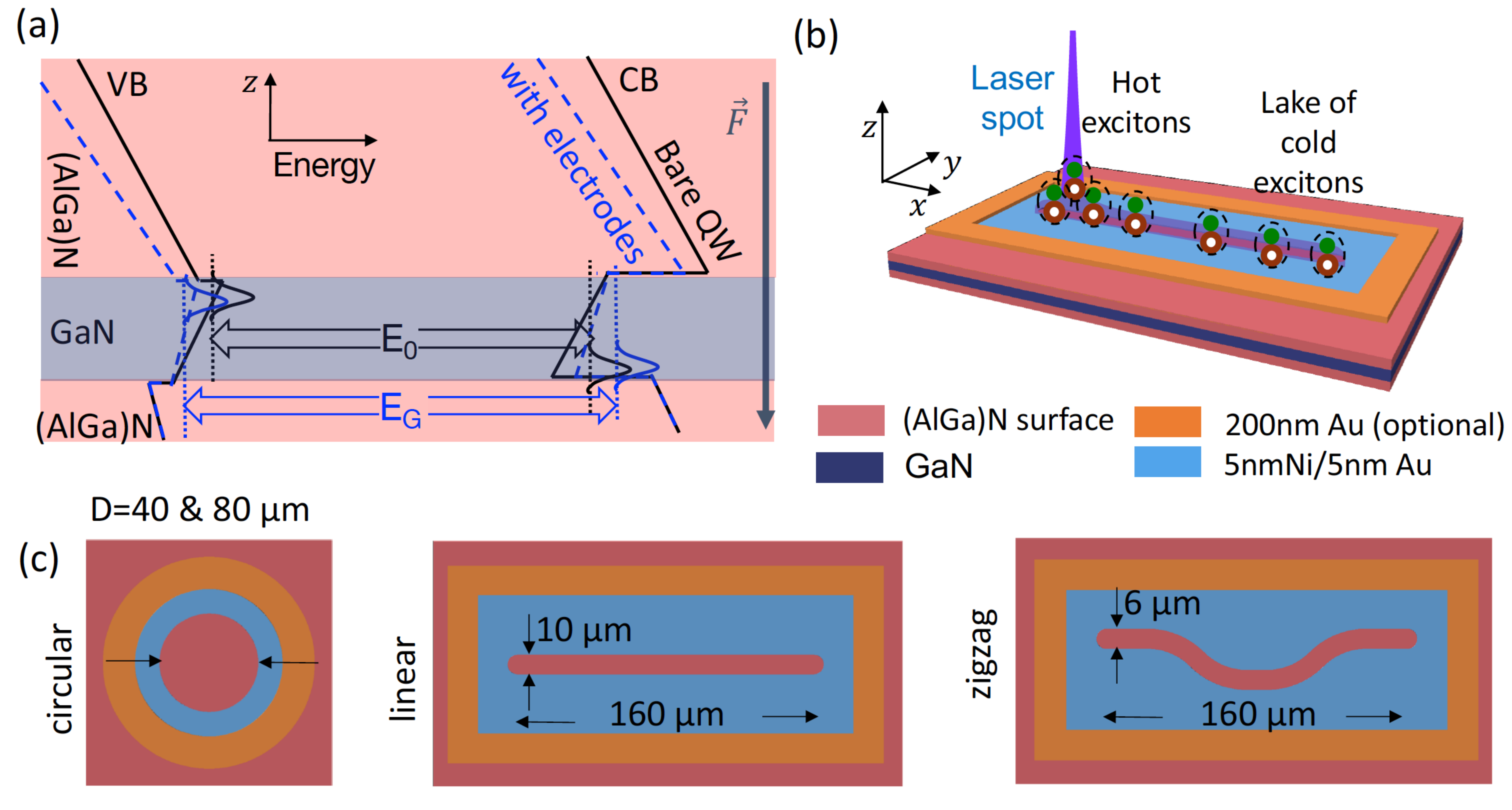}
  \caption{ (a) Band diagram of the GaN/(AlGa)N QW  in the absence (black solid line)  and in the presence of the top electrodes (blue dashed line). Corresponding carrier wavefunctions, energy levels, and interband transition energies are shown with the same colour code. (b) Artist view of the studied structure. (c) The three types of  traps under study.}
  \label{fig:fig1}
\end{figure}

Indeed, in GaN/(AlGa)N heterostructures the built-in electric field due to spontaneous and piezoelectric polarization 
is of order of megavolt per centimeter \cite{Bernardini1997,Leroux1998,Grandjean1999}. 
It pushes electrons and holes towards the two opposite interfaces of the QW. 
%This reduces the exciton binding energy below GaN bulk exciton ($\approx 25$~meV), but it remains still higher than maximum of $4$~meV reported for dipolar excitons in GaAs.
An exciton consisting of an electron and a hole spatially separated along the growth direction presents a permanent dipolar moment, roughly given by the QW width. 
The quantum confined Stark effect is so strong that the ground state exciton energy ($E_0$ in Fig. \ref{fig:fig1}~(a)) is lower (in this work by $\approx 300$ meV) than that of the exciton energy in bulk GaN.
The thin semitransparent metallic layer deposited on top of the structure reduces the electric field in the QW underneath, increasing exciton ground state energy: $E_G$  in Fig. \ref{fig:fig1}~(a). 
Using this effect,  the energy of the exciton can be modulated, making possible the on-demand patterning of the in-plane potential for the QW excitons.
Note that this trapping scheme is opposite to the one previously developed in GaAs QWs, where excitons are trapped in the regions covered by the metal, where the electric field is the strongest.

We also compare various sizes and shapes of the electrodes and show that quasi-one-dimensional linear traps are optimal host structures for the exciton fluids (Fig. \ref{fig:fig1}~(b)).
 They allow us to avoid the dilution of the exciton fluid due to radial extension, and thus to achieve high exciton densities as far as $100$~$\mu$m away from the excitation spot. Moreover, in this geometry, the relative contribution of the parasitic guided-and-scattered light (which is not confined by the electrodes so that its intensity decreases as $1/r$, where $r$ is the distance from the excitation spot) is substantially reduced with respect to the confined excitons emission.  Finally, implementation of quasi-one-dimensional zigzag-shaped potential (Fig. \ref{fig:fig1}~(c)) provides unambiguous evidence of the diffusive transport of the trapped excitons.
\begin{figure}
\includegraphics [width=1\columnwidth] {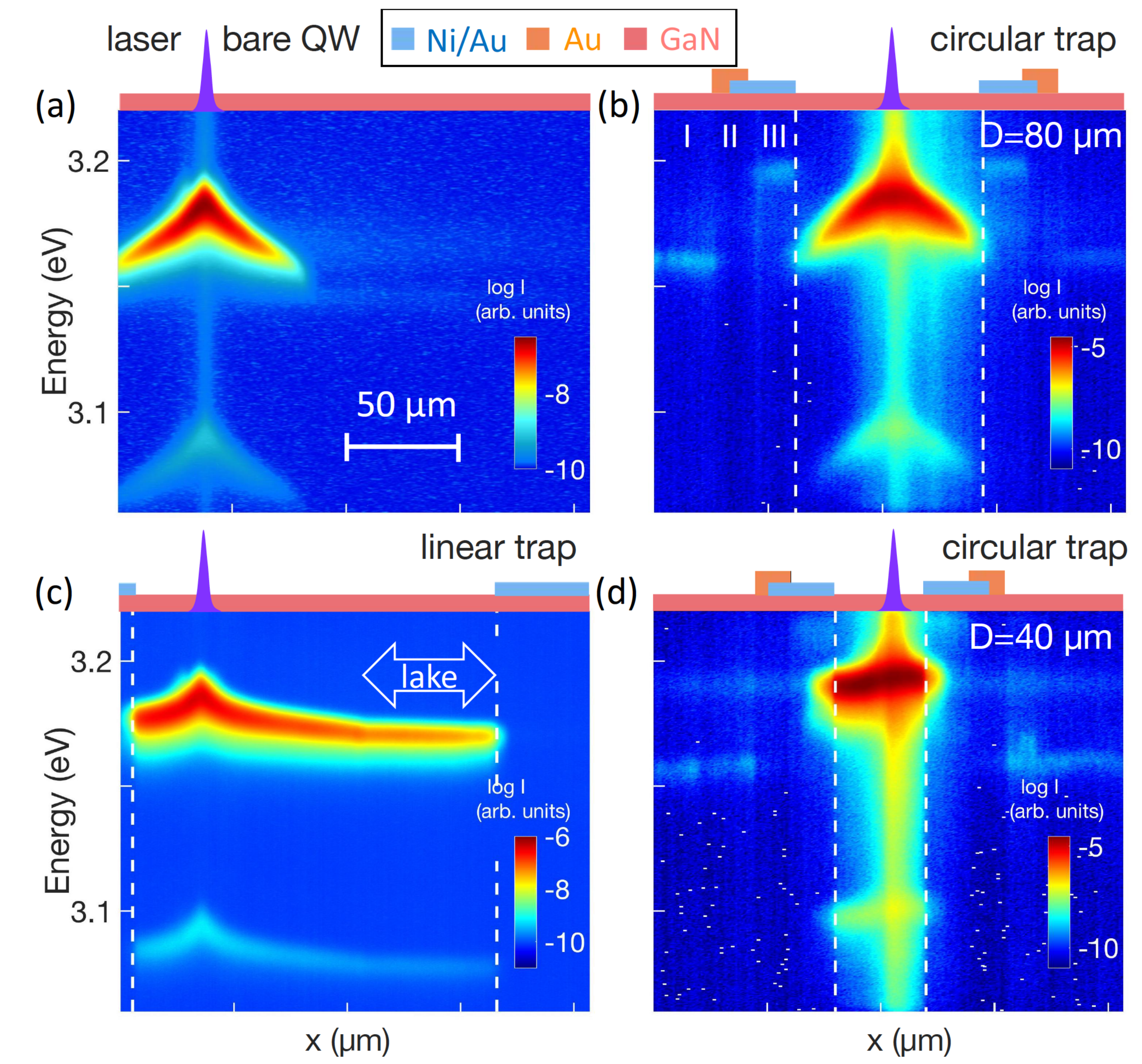}
  \caption{PL intensity (color-encoded in log scale) measured at $T=4$~K under point-like excitation of a bare QW (a) and of the QW patterned with electrodes providing circular (b), (d) and  linear (c) traps for the excitons. The electrode patterns and the laser spot position are shown schematically. Excitation powers are $P=20$~$\mu$W (a), $P=10$~$\mu$W (b, d) and $P=17$~$\mu$W (c).}
  \label{fig:bcl}
\end{figure}

%\section{Results and discussion}
We investigate a $7.8$~nm-wide GaN QW sandwiched between  
$100$ (top) and $50$~nm-wide (bottom) Al${_0.11}$Ga$_{0.89}$N barriers, grown on a free-standing GaN substrate (LUMILOG, threading dislocation density $2\times10^7$~cm$^{-2}$ ). 
The exciton radiative lifetime in such structure (in the zero zero-density limit) is estimated as $\tau_0 \approx 10$~$\mu$s \cite{Fedichkin2016}, 
exciton binding energy as $E_b=10$~meV and the built-in electric field $F=779$~kV/cm.
The semi-transparent electrodes consisting of $5$~nm of Au on top of $5$~nm of Ni are evaporated on the sample surface and patterned by optical lithography. 
Some of the patterns are additionally covered by a $200$~nm-thick Au layer  for subsequent wire bonding (Fig. ~\ref{fig:fig1}~(b)). This layer is opaque, but is not expected not affect the surface potential created by the NiAu electrodes. 
Three electrode geometries used to create in-plane trapping potential are illustrated in Fig. ~\ref{fig:fig1}~(c): circular trap, linear trap, and zigzag-shaped trap. 
In contrast with GaAs-based devices, the regions with bare surface are those where the electric field is the strongest,  and where excitons accumulate due to maximised confinement, while under the electrodes the field is lower. 
The details on the calculation of the band profiles and the confinement energies using coupled Schr\"{o}﻿﻿dinger and Poisson equations are provided in Supporting information. These calculations yield a trapping potential $E_G-E_0= 30$~meV, in excellent agreement with experimental results, as shown below. 
Here $E_0$ and $E_G$ are the exciton energies in the bare QW and in the QW covered by the electrodes, respectively, see Fig. ~\ref{fig:fig1}~(a).

In most of  our experiments the sample is cooled down to $4$~K and the QW excitons are created within a $1$~$\mu$m-diameter spot on the sample surface. We use either continuous wave (CW) optical pumping at $\lambda=266$~nm, or quasi-CW pumping at $360$~nm (picosecond pulses at $80$~MHz frequency, much higher than inverse exciton lifetime $1/\tau_0 \approx 0.1$~MHz) .  The PL spectra are acquired  by a charge coupled device (CCD) camera with $\approx 1$~$\mu$m spatial resolution. More details can be found in \citet{Fedichkin2016} and in Supporting information.

The exciton transport and trapping, as well as their dependence on the trap geometry, are presented in Fig. \ref{fig:bcl}.
It shows color-encoded  PL  spectra, measured at various distances from the excitation spot along the $x$-axis ({\it cf} Fig.~\ref{fig:fig1}~(b)). 
The emission from the bare surface (Fig. \ref{fig:bcl}~(a), no trapping electrodes) is compared to the circular traps with diameters $D=80$~$\mu$m and $D=40$~$\mu$m (Fig. \ref{fig:bcl}~(b), (d), circular traps) and quasi-one-dimensional channel ( Fig. \ref{fig:bcl}~(c), linear trap). 
Apart from the broad emission directly under the excitation point, at any distance from the spot the spectrum exhibits two main peaks: the zero-phonon exciton emission line and the first phonon replica $91$~meV below \cite{Zhang2001}.
The energy, intensity, and linewidth of the excitonic PL decrease when moving away from the excitation position. 
These effects are due to the screening of the electric field along the growth direction induced by the photocreated carriers: the highest density under the pumping spot corresponds to the maximum screening  (lowest electric field), and thus to the highest emission intensity and energy \cite{Lefebvre2004,Fedichkin2015,Fedichkin2016}. 
The pump-induced emission energy shift is referred to as the blue shift, $E_{BS}$. Its magnitude  roughly corresponds to the repulsive interaction between the dipolar excitons.
$E_{BS}$ can be used for estimating the exciton density as $n=E_{BS}/\phi_0$, where 
$\phi_0=1.5\times10^{-13}$~eV cm$^{-2}$ is the coefficient extracted from the self-consistent 
solution of the Schr\"{o}﻿﻿dinger and Poisson equations (see also Supporting information) \cite{Lefebvre2004,Fedichkin2015,Fedichkin2016}. 
This estimate could be improved taking into account excitonic correlations (see discussion below) \cite{Zimmermann,Laikhtman,Laikhtman2009}. 
%
%The exciton linewidth increases with increasing density \cite{Honold1989,Rossbach2014}. COMMENT
%
\begin{figure}
\includegraphics [width=1\columnwidth] {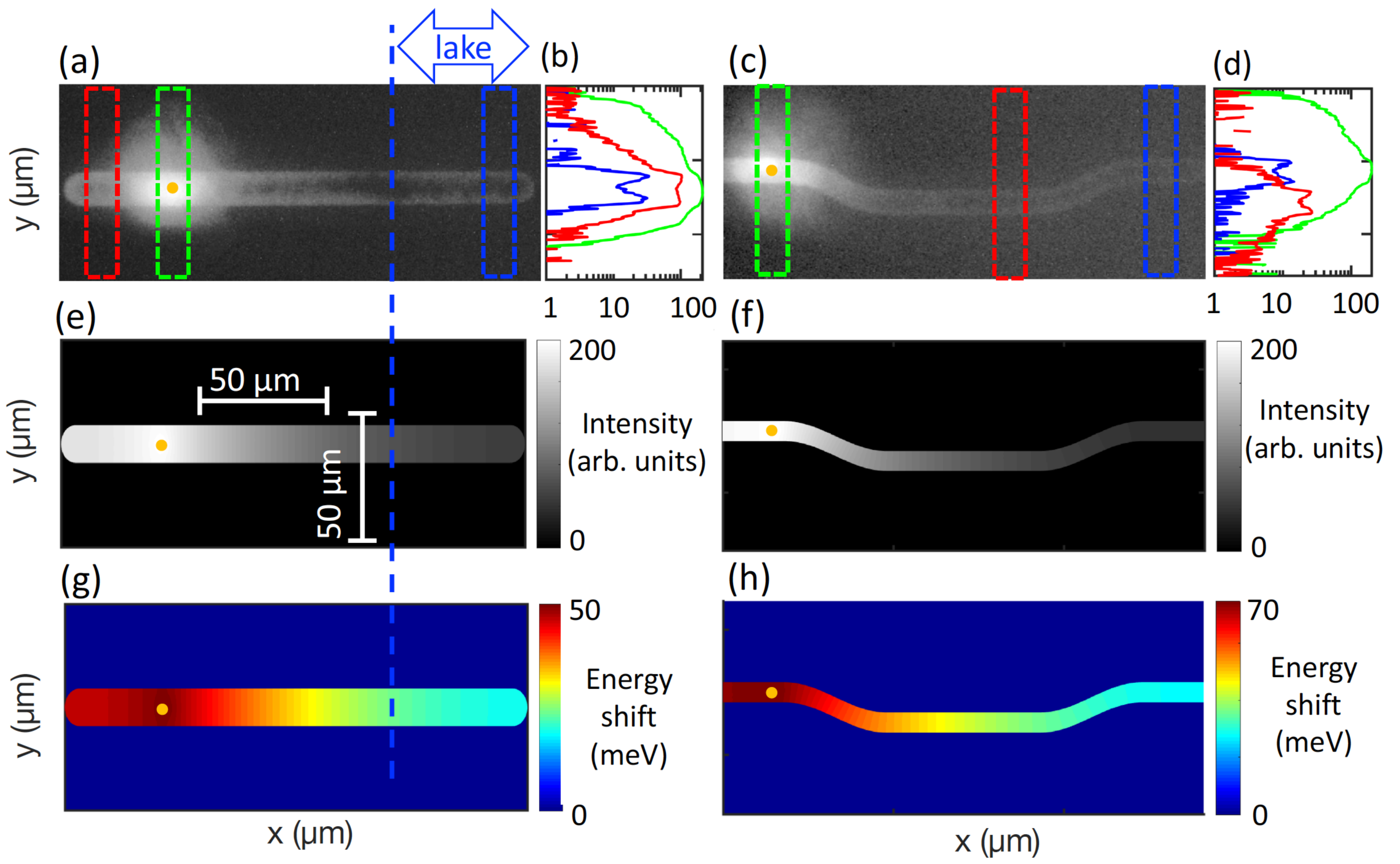}
  \caption{Grey scale-encoded $xy$ map of the spectrally-integrated ($3.05-3.25$~eV) PL  intensity in the linear (a) and zigzag-shaped (c) traps at $P=1$~mW. Yellow dots indicate the laser spot position.
  The intensity profiles shown (b) for linear trap and (d) for zigzag trap correspond to intensities integrated over $x$, in the windows displayed by dashed lines of the same color.
%   Dashed frames show the spatial areas corresponding to the integrated-over-x profiles in (b) for the linear and in (d) for the zigzag trap.
  %
 (e-h) are the results of the  drift-diffusion modelling. Both intensity (e-f) and energy shift $E_{BS}$ (g-h) are shown. Blue dashed lines delimit the part of the linear trap denoted as lake, where the variations of the PL intensity and energy (and thus exciton density) are weak. The intensity scales in (a), (c) and  (e-f) are identical. }
  \label{fig:transport}
\end{figure}

Let us compare the spatial profiles of the emission from the bare QW and circular traps. 
In the bare sample an arrow-shaped pattern with $\approx 40$~$\mu$m-width is accompanied by  weak emission far away from the spot at an approximately constant energy of $3.145$~eV, as previously observed in similar samples \cite{Fedichkin2016}.
In the sample with $80$~$\mu$m circular trap  an arrow-shaped pattern of similar size can be identified. 
This means that we load the trap with excitons, but the density across the trap  is not homogeneous.  
By contrast, in a smaller trap, the arrow-shaped pattern is replaced by a flat pattern. 
This suggests  that the in-plane potential is completely  screened by the trapped excitons, or, in other words, that the trap is fully filled. 

In both circular traps we observe weak, but clearly patterned emission from the outer regions of the trap.
Three different zones can by distinguished outside the trap: bare QW (I), thick gold electrode (II) and semitransparent electrode (III). 
The emission from the zone I is centered at the energy $E_0$ of the empty QW.
The excitons that emit this light have escaped from the trap and propagated diffusively, their density is small enough to produce no measurable blue shift.
This emission is alike the emission in a bare sample  more than $60$~$\mu$m away from the spot.
The small difference in the energy between the bare sample ($\approx 3.15$~eV in panel (a)) and the "outside the trap" zone I ($\approx 3.16$~eV in panels (b) and (d)) could be due to electrostatic effects induced by closely lying electrodes. 
In the smaller circular trap one can also see in zone I an emission at the same energy as the strongest emission at the center of the trap ($\approx 3.18$~eV). 
It is alike broad emission in bare sample at more than $60$~$\mu$m from the spot. 
This light is emitted by the excitons in the  trap, guided along the sample plane over tens of microns, and then scattered out of  the sample by surface roughness.
Indeed, this emission is also present in zone II, where opaque gold electrode impedes any emission that originates from the QW, independently on the size of the trap.
Finally, the emission in zone III, the closest to the trap,  provides an estimation of the trap depth.
This is particularly clear for the $80$~$\mu$m-wide trap.  A few excitons that escaped from the trap in the region covered by the semitransparent electrodes emit light at  $E_G=3.19$~eV. This corresponds to the modulation of the confinement potential induced by the metallic gates. Remarkably,  the measured trapping potential $E_t=E_G-E_0\sim 30$~meV matches perfectly the value expected from the model based on the solution of the coupled Schr\"{o}﻿﻿dinger and Poisson equations (see Supporting Information).

The advantage of the linear trap with respect to the circular one is quite natural: it allows us to eliminate the radial dilution of the exciton density ($\propto 1/$distance), and thus  to keep relatively high densities of excitons at significant distances from the excitation spot. 
Moreover, in circular traps, as well as  in the bare QW, the exciton emission competes with the guided-and-scattered light (although these two kinds of emissions can be spectrally distinguished).
By contrast, in the linear traps the guided-and-scattered light is subject to the radial dilution, while the excitons are confined in the trap, which is quasi-one-dimensional.

The linear trap (Fig. \ref{fig:bcl}~(c)) exhibits a particularly remarkable emission pattern. 
While close to the excitation spot one can still distinguish an arrow-shaped pattern, starting from $\approx 50$~$\mu$m and further away from the excitation spot the emission energy  and intensity are almost independent of the position. 
This suggests that we have created, within the trap, an area homogeneously filled with carriers, that will be referred to as an exciton lake. 
At $P=17$~$\mu$W used in Fig. \ref{fig:bcl}~(c), the blue shift of the exciton energy in the lake  does not exceed $20$~meV. 
This is less than the trapping potential, making possible the accumulation of the excitons in the trap.

More insight into exciton in-plane propagation and trapping can be obtained from the spectrally integrated emission intensity maps shown in Fig. \ref{fig:transport}~(a) and (c) for linear and zigzag-shaped traps, respectively. 
These images are acquired through a CMOS camera after a spectral filter matching the energy range shown in  Fig. \ref{fig:bcl}. 
Two main features can be distinguished in these images. 
First, along the $y$-direction the emission from the traps is inhomogeneous.
The emission from the borders is stronger than the one from the center of the trap. 
This is additionally illustrated in Fig. \ref{fig:transport}~(b) and (d), where we show the profiles taken under the spot (green line) and at two other positions along the $x$-direction (red and blue lines) for linear and zigzag-shaped traps, respectively. 
The observed inhomogeneity is probably related to the shape of the trap, suggesting that the electric field in $z$-direction varies progressively between the in-plane regions covered by the metal and the regions with bare  surface. 

Second, under the pumping power used in this particular experiment ($~1$mW)  the traps are strongly leaking. 
Indeed, the laser spot  (position indicated by the yellow dot) is surrounded by a circular halo, 
suggesting that under the spot the carrier density is high enough to screen the trapping potential
$E_t\approx 30$~meV. 
This halo being radially symmetric, it may include contribution from the photon diffusion, analogous to the one in Fig. \ref{fig:bcl}~(a). 
However, the in-plane profiles of the emission in the remote regions of the traps are inconsistent with the radial symmetry of the light propagation. 
This is particularly obvious from  the zigzag-shaped trap emission. 
Thus, these experiments prove unambiguously that the emission as far as $100$~$\mu$m away from the excitation spot results from the exciton transport, rather than from photon diffusion. 

To go further and quantify relevant parameters of the exciton propagation, we apply the exciton transport model developed in \citet{Rapaport2006,Fedichkin2016} to the case of transport in a two-dimensional trapping potential.
For the sake of simplicity, we neglect some experimentally observed features: we assume infinite trap barriers (no leakage) and spatially homogeneous potential within the trap. 
The exciton transport is assumed
to be limited by the presence of randomly distributed scatterers (Model C of Ref. \cite{Fedichkin2016}) so that the diffusion coefficient $D$ is given by  $D=l\sqrt{2k_BT/M}$, where $M=1.2m_0$ is the exciton mass, $k_B$ is the Boltzmann constant, $m_0$ is the free electron mass, and $l=10$~nm is the exciton scattering length. The latter, together with the fraction of the photons absorbed in the QW $\alpha=10^{-4}$ are the only fitting parameters. 
The details on the transport modelling are given in the Supporting Information.
The results of the numerical solution of the diffusion equation are shown in Fig. \ref{fig:transport}~(e-h) for both linear (e, g) and zigzag-shaped (f, h) traps. 
The simulation reproduces the propagation and filling of the traps, as well as the formation of the region $\approx 100$~$\mu$m away from the laser spot, where exciton emission energy and intensity are almost constant: the exciton lake.
%
%Note, that the exciton scattering length 
%
\begin{figure}
\includegraphics [width=1\columnwidth] {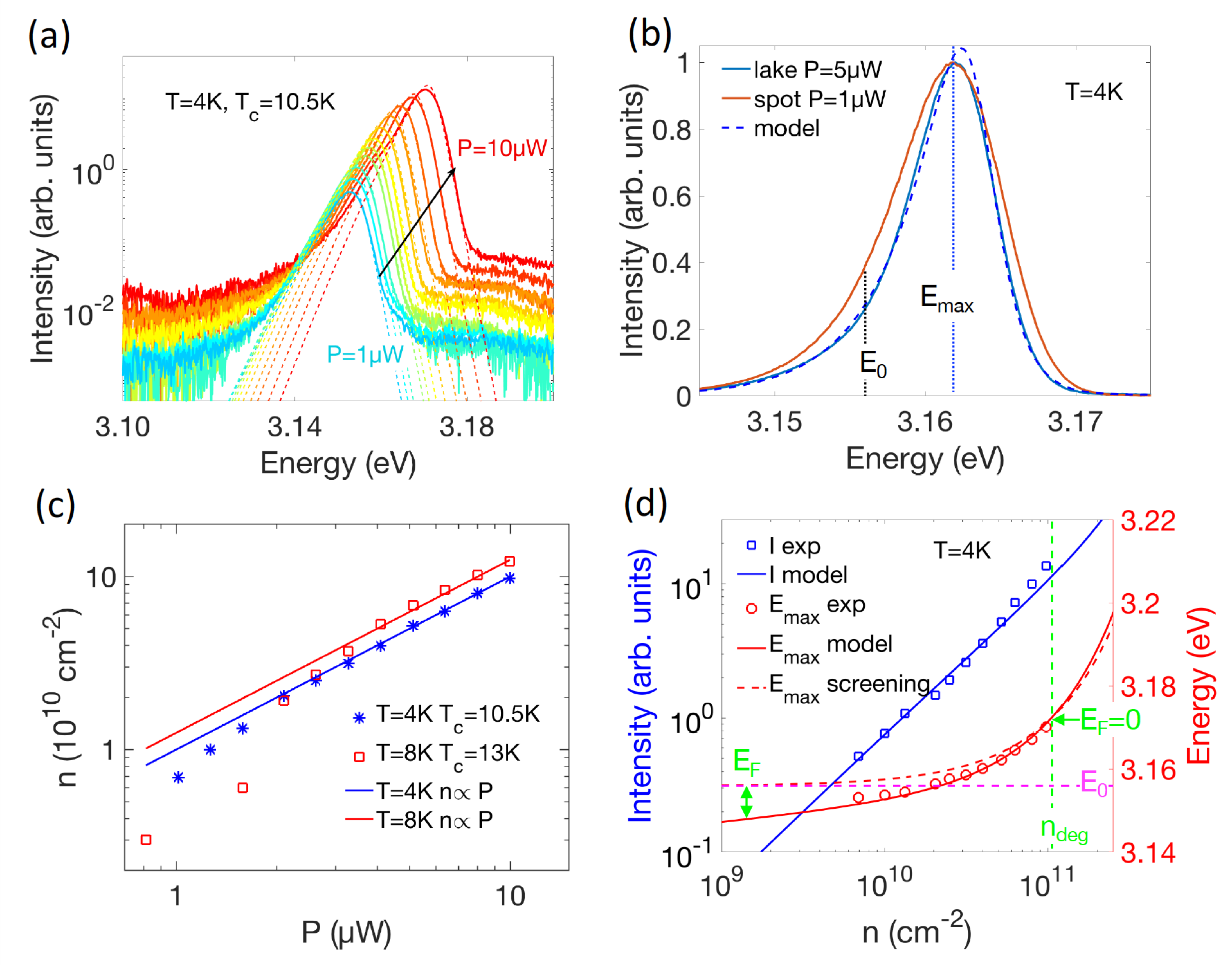}
  \caption{(a) Exciton emission spectra integrated over the lake area at different powers (solid lines) and the results of the fitting procedure with power-independent carrier temperature $T_c$ (dashed lines). (b) Comparison between the normalized spectra in the lake at $P=5$~$\mu$W and under the laser spot at $P=1$~$\mu$W. Vertical lines point the energies $E_0$ of the empty QW and $E_{max}$ corresponding to the  intensity peak at a given power. (c) Exciton density  as a function of the power extracted from the modelling of the spectra at $4$~K with $T_c=10.5$~K shown in (a) and  at $8$~K with $T_c=13.5$~K (spectra not shown). Solid lines are guides for the eyes indicating linear slope. (d) Integrated intensity (left scale) and the spectral peak energy $E_{max}$ (right scale) as a function of the carrier density extracted from the data (symbols) and the model (solid lines) shown in (a). The dashed lines show the density corresponding to the degeneracy condition (green), the energy of the empty  QW $E_0$ (magenta) and the calculated variation of the QW energy $E_{max}$ due to carrier-induced screening of the built-in electric field (red).}
  \label{fig:spectra}
\end{figure}

In the  following we analyze the spectral properties of the emission from the lake region in the linear trap as a function of excitation power density, in order to demonstrate density and temperature control of the dipolar excitons.
%
%The spectrum of the exciton fluid in the macroscopic homogeneously filled region of the trap -the lake- is now studied as a function of the exciton density.
%
Figure \ref{fig:spectra}~(a) shows a set of exciton emission spectra (solid lines) averaged over the lake region of the linear trap (white arrow in Fig. \ref{fig:bcl}~(c)).  The power varies from $1$ to $10$~$\mu$W and sample temperature is fixed at $T=4$~K.
One can identify a blue shift of the emission line ($\approx 20$~meV) and increasing intensity that accompanies the increase of  the power. 
The line shape is not Gaussian, as one could expect for the emission dominated by inhomogeneous broadening.
Instead, an exponential shape of both high- and low-energy tails is apparent.
Similar  profiles have already been observed in PL spectra of dipolar excitons \cite{Stern2014,Schinner2013}. 
In particular, Schinner {\it et al} use  Fermi-Dirac distribution as a cutoff function, to account for the thermal smearing of the filling of the trap with the dipolar exciton liquid \cite{Schinner2013}.
An argument that can be put forward in favor of this approach is that in the case of the bosonic dipolar liquid, the strong spatial dipolar repulsion plays  the role of the Pauli principle in the Fermi-Dirac distribution. 
To account for the inhomogeneous broadening  and describe the low-energy tail of the spectrum, Schinner  {\it et al} multiply the  Fermi-Dirac distribution by a phenomenological Gaussian function  \cite{Schinner2013}.
In our samples the low-energy tail of the excitonic emission is rather exponential, similar to \citet{Stern2014}  
Therefore we model this peculiar shape with the product of  the exponential function (to describe the low-energy tail of the spectrum) and the Fermi-Dirac distribution (to account for the temperature-broadened steep decrease of the high-energy tail of the spectrum)\footnote{We limit ourselves to the fundamental electronic transition involved in the A-exciton, because we do not observe any evidence of the distinct emission of neither excited states exciton states, nor B-exciton, the emission linewidth  $\approx 8$~meV being of the same order of magnitude as the typical splitting between A and B excitons.}.

The efficiency of the exciton cooling in the lake can be estimated by fitting this function to the set of data at different powers, as shown in Fig. \ref{fig:spectra}~(a). 
From such a procedure, we extract the following fitting parameters: the zero-density exciton energy $E_0=3.156$~eV, the carrier densities $n$, and the carrier temperature $T_c=10.5$~K.
The latter appears to be power independent,  albeit different from the nominal sample temperature $T=4$~K. 
%
%The values of $n$, $T_c$ and $E_0$ are the only fitting parameters used to describe the whole data set.
% 
The resulting carrier densities are shown in Fig. \ref{fig:spectra}~(c)  as a function of the laser power for two sets of data at different nominal temperatures. For $T=4$~K we get $T_c=10.5$~K and for $T=8$~K we get $T_c=13$~K. 
We speculate that the difference between carrier temperature and the nominal lattice temperature results from the poor thermalization of the 
crystal lattice in the cold finger cryostat, and the real lattice temperature is higher, so that $T\approx T_c$.
Apart from the very lowest powers ($P<2$~$\mu$W, where nonradiative recombination may dominate over the radiative one), the carrier density increases linearly with power, 
indicating vanishing contribution of nonradiative losses. The higher density at higher temperature is a signature of the increasing exciton lifetime in the QW \cite{Rosales2013}.

It is instructive to compare the emission peak with a given energy measured in the lake ($P=5$~$\mu$W), and directly at the excitation spot, but at lower  power, $P=1$~$\mu$W. 
Remarkably, even at a factor of $5$ lower excitation power, the emission under the spot is broader than the one in the lake and does not have the same profile,  so that the fitting procedure used for the lake spectra is not viable.
This suggests that the impact of optical excitation goes beyond the heating of the exciton gas and the screening of the built-in electric field, that are both taken into account by the model.

The determination of the carrier density in the nanostructures hosting dipolar excitons is a very delicate question since it relies on the knowledge of the exciton-exciton interactions. 
The values of $n$ given in Fig. {\ref{fig:spectra}~(c-d) are obtained in the rough semi-phenomenological approximation described above, neglecting 
excitonic correlations: $n=E_{BS}/\phi_0$, where $\phi_0$ is deduced from the self-consistent solution of the Schr\"{o}﻿﻿dinger and Poisson equations. 
The same approximation was used in  Refs. \cite{Fedichkin2015,Fedichkin2016}, and it is conceptually similar to the plate capacitor model \cite{Butov2001,Ivanov2002}. 
The effect of the exciton-exciton interactions was 
investigated theoretically and experimentally in Refs. \cite{Zimmermann,Laikhtman,Voros2009,Remeika2009,Remeika2015}  and results in a density
and temperature-dependent correction factor that leads to the underestimation of the density 
by up to one order of magnitude at the lowest powers and temperatures.
Nevertheless, it is not straightforward to apply those considerations to our system, and a 
more advanced analysis of the data remains to be done. 
In this work we determine the lower limit for the carrier density, and bear in mind that excitonic correlations may increase it by up to a factor of  $10$.

Fig. \ref{fig:spectra}~(d) summarizes the results of the modelling.
Both energy (red circles, right scale) and integrated intensity (blue circles, left scale) of the excitonic emission are shown as a function of the exciton density extracted from the model (solid lines). 
This representation allows us to appreciate the role of the two density-induced contributions to the shift of the emission energy:  the screening of the built-in electric field ($E_{max}=E_0+n\phi_0$, red dashed line)  and the phase space filling that leads to an additional shift of the emission energy. 
The latter can be described in terms of the variation of the Fermi energy $E_F$, that appears in  Fig. \ref{fig:spectra}~(d) as the difference between the red solid line, showing the full model and red dashed line, showing the screening effect on the emission energy. 
One can see that the Fermi contribution is not negligible at low densities, at least within the approximations described above.
Importantly, even at maximum carrier densities achieved in the lake the carriers remain in the non-degenerate limit $n<n_{deg}\approx 10^{11}$~cm$^{-2}$ (  so that $E_F<0$) and well below exitonic Mott transition expected to occur progressively above  $n_{Mott}=6\times10^{11}$~cm$^{-2}$ \cite{Rossbach2014}.
Moreover, the integrated emission intensity is well described by the model, and it increases almost linearly with the exciton density, at least up to $P=5$~$\mu$W. 
This means that here the carrier densities are low enough, so that exciton radiative emission rate does not grow exponentially with density,  as it is usually observed in polar heterostructures \cite{Lefebvre2004,Fedichkin2015,Fedichkin2016,Kuznetsova2015,Liu2016,Chen2018}. 

In conclusion, we have shown that dipolar excitons can be efficiently trapped in the plane of GaN/(AlGa)N QWs. 
In previous works on GaN QWs hosting dipolar excitons, the mutual repulsion of excitons led to a fast radial  expansion and dilution of the exciton gas, accompanied by dramatic variation of the exciton energy and lifetime \cite{Fedichkin2015,Fedichkin2016}.
Here we overcome these problems, and create spatial areas (referred to as lakes) with tens of micrometers characteristic size, homogeneously filled with excitons with well-defined emission energy and lifetime.
We show that the exciton fluid trapped in the lake is at macroscopic thermodynamical equilibrium with well-defined temperature close to the sample temperature. 
The density of this cold exciton fluid can be controlled by changing the laser excitation power.

The realisation of traps relies on the deposition of metallic gates on the surface.
Both the electrode shape, and the characteristic electric fields that must be manipulated (of order of $F=1$~MV/cm) are quite different from the known GaAs-based nanostructures.
Such strong built-in electric field in nitride-based structures could be not only a challenge to meet, but also an opportunity to avoid spurious photocurrents that limit exciton mobility and coherence properties.
These photocurrents are not easy to avoid in other materials hosting dipolar excitons (either GaAs-based or Van der Waals heterostructures), where a voltage must be applied across the regions where excitons accumulate.
High exciton binding energy in wide band gap semiconductors, such as GaN and ZnO, make them potentially promising for the realisation of exciton-based optoelectronic devices operating at room temperature.
Our results pave the way towards such functionalities, although the electrical control of the exciton fluxes remains to be demonstrated.
Finally, the detailed spectroscopic study of dipolar exciton fluids at various densities made possible by the presented results, may shed new light on their intriguing properties, such as macroscopic coherence and condensation, predicted to show up in GaN QWs at higher temperatures than in GaAs.

\begin{acknowledgement}

The authors thanks  D. Scalbert for enlightening discussions.
This work was supported by  the French National Research Agency (ANR OBELIX). 

\end{acknowledgement}

\begin{suppinfo}

Experimental methods; Modelling of the density-dependent confinement potential and carrier wavefunctions by self-consistent solution of the Schr\"{o}﻿﻿dinger and Poisson equations; Drift-diffusion modelling for the exciton transport; Modelling of the PL spectra, estimation of the carrier density.

\end{suppinfo}

\bibliography{biblio}

\end{document}